\documentclass[conference]{IEEEtran}
\IEEEoverridecommandlockouts

\usepackage{tablefootnote}
\usepackage{algorithmic}
\usepackage{graphicx}
\usepackage{textcomp}
\usepackage[table]{xcolor}
\usepackage{minted}
\usepackage{hyperref}
\usepackage{multirow}
\usepackage{enumitem}           
\usepackage[T1]{fontenc} 
\usepackage{tikz} 
\usepackage{stfloats}
\usepackage[utf8]{inputenc}
\usepackage{soul}
\usepackage{booktabs}
\usepackage{multirow}
\usepackage{colortbl}
\definecolor{mygray}{rgb}{0.71,0.71,0.71}
\definecolor{codegray}{rgb}{0.95,0.95,0.95}
\definecolor{codegreen}{rgb}{0,0.5,0}
\definecolor{codepurple}{rgb}{0.4,0,0.6}
\definecolor{dangerred}{rgb}{0.75,0,0}
\definecolor{safegreen}{rgb}{0,0.55,0}
\usetikzlibrary{shapes.geometric, positioning, arrows.meta}

\usepackage{balance}
\usepackage{amsfonts}
\usepackage[caption=false]{subfig} 
\usepackage{array}
\usepackage{cite}
\newcolumntype{P}[1]{>{\centering\arraybackslash}p{#1}}
\newcommand{\figref}[1]{Fig.~\ref{#1}}

\newcommand{\ie}{\textit{i.e.},\ }
\newcommand{\eg}{\textit{e.g.},\ }

\newcommand{\etc}{{\em etc.}}

\usepackage{booktabs}

\definecolor{francBlue}{RGB}{64,76,87}

\usepackage[most]{tcolorbox}
\usepackage[parfill]{parskip}

\tcbset{
  parskip/.style={
    before={\par\pagebreak[0]\parindent=0pt},
    after={\parfillskip=0pt\par},
  },
}
\newtcolorbox{resultbox}[1][]{%
    colback=black!3,
    colframe=black!3,
    notitle,
    sharp corners,
    borderline west={2pt}{0pt}{gray!80!black},
    enhanced,
    breakable,
    boxsep=0pt,
    left=4pt,right=2pt,top=2pt,bottom=2pt,
    }
\newcommand{\rques}[2]{  
\begin{itemize}[leftmargin=23pt, itemsep=0pt, parsep=0pt, topsep=0pt, partopsep=0pt, labelsep=0.4em]
\item[\textbf{RQ\textsubscript{#1}}] \textbf{#2}
\end{itemize}
}

\definecolor{codebg}{rgb}{0.99,0.99,0.99}
\definecolor{hiliteColor}{rgb}{1,0.92549019607,0.6}
\definecolor{tainted}{rgb}{0,1,1}

\newmintinline{python}{fontsize=\normalsize}

\newminted[PythonSourceCode]{python}{
    labelposition=topline,
    frame=single,
    numbersep=2pt,
    fontsize=\tiny,
    xleftmargin=5pt,
    framerule=0.5pt,
    linenos
}

\newminted[JavaSourceCode]{java}{
    labelposition=topline,
    frame=single,
    numbersep=2pt,
    fontsize=\tiny,
    xleftmargin=5pt,
    framerule=0.5pt,
    linenos
}

\definecolor{magnolia}{rgb}{0.97, 0.96, 1.0}
\definecolor{shadecolor}{rgb}{0.97, 0.96, 1.0}

\newmintinline[codePython]{python}{fontsize=\small}
\newmintinline[codeJava]{java}{fontsize=\small}
\newmintinline[snippetJava]{java}{fontsize=\small,bgcolor=magnolia}
\newmintinline[snippetPython]{python}{fontsize=\small,bgcolor=magnolia}
\newmintinline[snippetPerl]{perl}{fontsize=\small,bgcolor=magnolia}

\newcommand{\joanna}[1]{\textcolor{magenta}{}}

\newcommand{\subparagraph}[1]{\noindent\textbf{#1}\quad}

\newcommand{\ece}{\mathrm{ECE}}
\newcommand{\brier}{\mathrm{Brier}}
\newcommand{\ft}{\mathrm{FT}}

\newcommand{\bg}{\mathrm{OG}}
\newcommand{\br}{\mathrm{BS}}
\newcommand{\psec}{p^{\mathrm{sec}}}
\newcommand{\pfunc}{p^{\mathrm{func}}}

\newcommand{\highlightred}[1]{\textcolor{dangerred}{\textbf{#1}}}
\newcommand{\highlightgreen}[1]{\textcolor{safegreen}{\textbf{#1}}}

\definecolor{codegreen}{rgb}{0,0.6,0}
\definecolor{codegray}{rgb}{0.5,0.5,0.5}
\definecolor{codepurple}{rgb}{0.58,0,0.82}
\definecolor{backcolour}{rgb}{0.95,0.95,0.92}

\lstdefinestyle{mystyle}{
    backgroundcolor=\color{backcolour},   
    commentstyle=\color{codegreen},
    keywordstyle=\color{magenta},
    numberstyle=\tiny\color{codegray},
    stringstyle=\color{codepurple},
    basicstyle=\ttfamily\footnotesize,
    breakatwhitespace=false,         
    breaklines=true,                 
    captionpos=b,                    
    keepspaces=true,                 
    numbers=left,                    
    numbersep=5pt,                  
    showspaces=false,                
    showstringspaces=false,
    showtabs=false,                  
    tabsize=2
}

\AtBeginDocument{%
  }

\begin{document}

\title{An Empirical Study of Security Calibration in Large Language Models for Code}

 \author{
 \IEEEauthorblockN{Mohammed Latif Siddiq}
 \IEEEauthorblockA{\textit{University of Notre Dame}\\
Notre Dame, IN, USA \\
msiddiq3@nd.edu}
\and
 \IEEEauthorblockN{Md. Nafiu Rahman}
 \IEEEauthorblockA{\textit{Brac University}\\
  Dhaka, Bangladesh\\
 nafiu.rahman@bracu.ac.bd}
  \and
 \IEEEauthorblockN{Joanna C. S. Santos}
 \IEEEauthorblockA{\textit{University of Notre Dame}\\
  Notre Dame, IN, USA \\
 joannacss@nd.edu}
}

\maketitle

\begin{abstract}
Large Language Models (LLMs) are rapidly transforming software development, yet their use in security-critical contexts raises a key question: do models know when their generated code is insecure? This property, known as \emph{calibration}, measures whether a model’s confidence aligns with the true correctness of its outputs. We present the first large-scale empirical study of \emph{security calibration} in LLM-generated code. We evaluate GPT-4o-mini, Gemini-2.0-Flash, and Qwen3-Coder-Next across multiple temperature settings on two complementary benchmarks: self-contained security tasks and multi-language repository-level contexts. Our results suggest that overconfidence is prevalent across the evaluated LLMs. Functional calibration is consistently worse than security calibration, suggesting that models estimate security outcomes more reliably than functional correctness, potentially because functional correctness depends on complex execution behavior. We also examine whether calibration-guided automated repair can help remediate vulnerabilities in LLM-generated code, finding only limited improvements while frequently introducing functional regressions. 
Moreover, we study different mitigation strategies for reducing False Trust, where models assign high confidence to vulnerable code. The results show that although architectural gating improves calibration on controlled benchmarks, calibration deteriorates in realistic repository-level settings, increasing the risk of high-confidence vulnerable outputs.

\looseness=-1


\end{abstract}

\begin{IEEEkeywords}
Large Language Models, Code Generation, Security Calibration, Vulnerability Detection, CWE, Model Confidence, False Trust
\end{IEEEkeywords}

\section{Introduction}
\label{sec:intro}

Large Language Models (LLMs) for code are increasingly integrated into modern software development environments, assisting developers with tasks such as boilerplate code generation, bug fixing, and refactoring \cite{siddiq2025security}. As these systems become embedded in developer workflows, they act as programming assistants that influence how software systems are maintained and evolved. This widespread adoption raises important  questions about the \textit{security implications of AI-generated code} in real-world development settings \cite{siddiq2022empirical}.  

Unlike \textit{functional correctness}, where failures manifest as test failures or runtime errors, \textit{vulnerabilities} may remain latent until exploitation. Prior work shows that LLMs can generate vulnerable code, including SQL injection, cross-site scripting (XSS), and insecure deserialization~\cite{pearce2022asleep,perry2023insecurecodeaiassistant,sandoval2023lost,siddiq2024sallm}. Even more concerning, models can produce such code with high confidence, encouraging developers to place false trust in the generated outputs \cite{lin2022truthfulqa,calibrationllm2025}. In practice, developers must quickly evaluate AI-generated suggestions and decide which ones require scrutiny. High-confidence outputs are more likely to be accepted without careful review~\cite{sandoval2023lost}, creating a systemic risk: vulnerable AI-generated code may propagate into production systems through misplaced trust in overconfident models.

Existing benchmarks such as HumanEval~\cite{chen2021codex}, MBPP~\cite{austin2021program}, and SWE-Bench~\cite{jimenez2024swebench} primarily evaluate \emph{functional correctness} using unit-test pass rates and provide little assessment of \emph{security properties}. Recent security-oriented benchmarks \cite{siddiq2024sallm,aicgseceval2025} evaluate whether generated code contains vulnerabilities, but do not measure whether models \emph{recognize} the security risks in their own outputs. That is, they measure whether \emph{the code is vulnerable}, but not whether \emph{models know, in the sense of accurately self-assessing, that it is vulnerable}. We use the term \emph{knowing} to denote \emph{probabilistic self-assessment of security correctness}: the degree to which a model's stated confidence in the security of its own output aligns with the output's actual security status, as determined by executable test outcomes. This notion is distinct from \emph{vulnerability reasoning}, in which a model detects or classifies known weakness patterns~\cite{siddiq2024sallm,aicgseceval2025}, and from \emph{self-explanation}, in which a model verbalizes its rationale without quantifying correctness likelihood~\cite{ullah2024secllmholmes}.

In machine learning, \textit{\textbf{calibration}} measures the alignment between a model’s confidence and its empirical accuracy. A well-calibrated model reporting ``80\% confidence'' should be correct  80\% of the time. Calibration is therefore critical for trustworthy AI: even accurate models become unreliable when their confidence estimates are systematically misaligned \cite{calibrationllm2025}. In code generation, poor calibration manifests as the \textit{\textbf{False Trust}} problem \cite{lofstrom2023definition}, where models assign high confidence to vulnerable code. Developers who rely on such estimates may deploy insecure code, creating a risk: missed vulnerabilities can cause severe damage (\eg data breaches or remote code execution), whereas false alarms only incur additional review cost. Consequently, we argue that \textbf{\textit{calibration}}, not just accuracy, should be a central evaluation criterion for LLMs in security-critical settings.

Despite growing research on vulnerabilities in LLM-generated code~\cite{siddiq2024sallm,meta2023cybereval,pearce2022asleep}, no prior work has systematically measured \emph{security calibration}. Recent work on LLM calibration in software engineering examines functional correctness~\cite{calibrationllm2025} or general detection rates~\cite{li2025safegenbench}, without considering security. Thus, four research gaps remain. First, existing studies report binary outcomes (secure \textit{vs.}\ insecure) without collecting model confidence, making calibration assessment difficult and leaving the relationship between functional calibration ($\pfunc$) and security calibration ($\psec$) unexplored. Second, it is unknown whether calibration patterns observed in self-contained programs generalize to realistic repository-level contexts with cross-file dependencies. Third, automated repair methods rarely leverage model uncertainty to guide vulnerability remediation, and it remains unclear whether calibration signals can drive effective repair. Fourth, mitigation strategies such as execution gating, prompt engineering, and few-shot prompting have not been evaluated for their impact on security calibration. 

To address these gaps, we present the first large-scale empirical study of \emph{security calibration} in LLM-generated code. We analyze whether models accurately estimate the security correctness of their outputs across thousands of samples, multiple models, and both controlled and repository-level development contexts. Our study provides new insights into how confidence signals from AI coding assistants can be interpreted, calibrated, and integrated into secure software development workflows. This work makes the following contributions\footnote{Data and scripts used in this work are available on our replication package: \url{https://github.com/s2e-lab/Calibrating-Secure-Code-Generation}}:
\begin{itemize}[leftmargin=*, itemsep=0pt, parsep=0pt, topsep=0pt, partopsep=0pt, labelsep=0.4em]

\item We conduct the first \textbf{\textit{large-scale study of security calibration in LLM models}} (RQ1).  
We investigate 36,000 generated code snippets across three models (GPT-4o-mini, Gemini-2.0-Flash, Qwen3-Coder-Next), six temperature settings, and 100 security-critical tasks. We measure Expected Calibration Error (ECE)~\cite{naeini2015obtaining}, Brier score~\cite{glenn1950verification}, False Trust rate, and overconfidence gap. Using these metrics, we compare \textit{security} and \textit{functional} confidence estimates to analyze differences between the two calibration signals.

\item 
We \textbf{\textit{quantify the calibration degradation when moving from self-contained programs to repository-level code generation across multiple programming languages}} (RQ2). Our study reveals consistent calibration penalties driven by cross-file structural complexity.

\item  
We \textbf{\textit{investigate whether a closed-loop repair system using post-hoc calibrated confidence thresholds is suitable for removing vulnerabilities in LLM-generated code}} (RQ3). Our study shows that, even after applying isotonic and logistic calibration, models rarely revise their outputs for vulnerabilities that require replacing an insecure API call or data sink with a secure alternative.

\item We \textbf{\textit{evaluate mitigation strategies for reducing False Trust in secure code generation}} (RQ4). We study three interventions: execution gating, authorized-breaking repair, and few-shot prompting (with and without chain-of-thought). Our results show that architectural interventions (execution gating) help improve calibration, while prompt-level techniques provide limited benefit under strict harness evaluation.
\end{itemize}
\section{Background and Related Work}\label{sec:background}

\subsection{LLMs for Code Generation}

LLMs are increasingly integrated into software development workflows through tools like GitHub Copilot. These systems translate natural language into code, generating boilerplate, API calls, and configuration artifacts \cite{chen2021codex,austin2021program,openai2024gpt4,li2023alphacode}. Since LLMs are trained on large corpora of open-source code, they learn both secure and insecure programming practices \cite{pearce2022asleep}. Prior works show that LLM-generated code may contain vulnerabilities  (\eg SQL injection, Cross-Site Scripting, unsafe deserialization, weak cryptographic uses,~\etc) \cite{pearce2022asleep,khoury2023how,perry2023insecurecodeaiassistant,siddiq2024securityeval}. These flaws often occur in syntactically valid and functionally correct programs, making them difficult to detect solely through functional testing.


Security correctness differs from functional correctness. \textit{Functional correctness} measures whether a program satisfies its intended input–output behavior and is commonly evaluated using benchmarks such as HumanEval and MBPP \cite{chen2021codex,austin2021program}. \textit{Security correctness}  concerns properties such as confidentiality, integrity, and availability under adversarial conditions. Vulnerabilities typically arise from unsafe data flows, using insecure APIs, or missing authorization checks \cite{mitrecwetop25,santos2017understanding}. 

To evaluate these risks,  benchmarks and frameworks have been introduced, including scenario-based vulnerability studies \cite{pearce2022asleep,debenedetti2023a}, security-oriented prompt suites  \cite{siddiq2024securityeval,siddiq2024re}, and automated pipelines combining prompts with static analysis  \cite{meta2023cybereval,siddiq2024sallm}. More recent frameworks (\eg AICGSecEval \cite{aicgseceval2025}) extend evaluation to multi-language and repository-level settings. However, these benchmarks typically report only binary outcomes (secure/insecure) and do not analyze the LLM's confidence in its outputs. Our work studies whether models are \emph{calibrated} about the security of the code they generate.

\subsection{Calibration and Confidence Estimation}

{\textit{Calibration}} measures the alignment between a model’s predicted confidence and empirical accuracy. A perfectly calibrated model satisfies
$P(Y' = Y \mid P(Y' \mid X) = p) = p$,
meaning that predictions made with confidence $p$ are correct approximately $p$ of the time \cite{jiangetal2021know}. Modern neural models, however, are often \emph{overconfident}, assigning high probability to incorrect predictions \cite{guo2017calibration}. In human–AI interaction, such miscalibration can produce the \textit{{False Trust}} phenomenon: users trust incorrect predictions because the system expresses high confidence \cite{lofstrom2023definition}. In code generation, this occurs when a model confidently produces insecure code, increasing the likelihood that developers accept vulnerable outputs.

Calibration is typically measured using metrics such as the Expected Calibration Error (ECE) and the Brier score \cite{glenn1950verification}, which are widely adopted in the literature \cite{naeini2015obtaining,nixon2019calibration}. Post-hoc calibration techniques, including \textit{temperature scaling }\cite{guo2017calibration}, \textit{Platt scaling} \cite{platt1999probabilistic}, and \textit{isotonic regression} \cite{niculescu2005predicting}, can partially correct miscalibration without retraining the model. In AI-assisted development, reliable calibration is important because developers may implicitly rely on model confidence when deciding whether to trust generated code.

Calibration requires estimating the LLM's \textit{confidence score}, which can be derived from several mechanisms. One approach \textit{aggregates token-level log probabilities} to approximate sequence likelihood~\cite{kadavath2022language}.  Another uses \textit{verbalized confidence}, prompting models to explicitly report their certainty~\cite{kadavath2022language,kuhn2023semantic}, often via specialized prompting or fine-tuning strategies~\cite{lin2022teaching,tianetal2023just}.  Confidence can also be inferred through \textit{sampling-based consistency}, where agreement across multiple outputs indicates reliability, or through reflection and self-consistency prompting that evaluates reasoning stability across sampled paths~\cite{wang2023selfconsistency,wei2022cotprompting}. Each method has trade-offs: token-based approaches require API access, sampling methods are computationally expensive, and verbalized confidence may reflect linguistic bias rather than true uncertainty. Importantly, most prior work evaluates confidence on general tasks or functional correctness rather than the security properties of generated code \cite{calibrationllm2025,kadavath2022language,desaidurrett2020calibration,jiangetal2021know}.

\subsection{Calibration for LLMs and Software Engineering}

Prior research shows that LLM confidence is often poorly aligned with empirical correctness across tasks such as question answering and reasoning \cite{desaidurrett2020calibration,jiangetal2021know}. While post-hoc methods can partially improve calibration, misalignment remains common. Some work suggests that LLMs can produce calibrated self-assessments with appropriate prompting \cite{kadavath2022language}, while other approaches estimate correctness probabilities through specification-based verification \cite{key2023trustworthyneuralprogramsynthesis}. However, these studies focus primarily on natural language tasks or general program synthesis rather than security-critical code generation.

Within software engineering, calibration remains underexplored. Recent studies show that confidence estimates derived from token probabilities or self-evaluation often misalign with functional correctness in code generation tasks \cite{zhou2024calibration,calibrationllm2025}. These analyses focus exclusively on functionality and do not examine calibration with respect to security properties. Thus, we fill this gap by conducting the first study of \textit{security calibration} in LLM-generated code, measuring the alignment between model confidence and vulnerability outcomes across both self-contained and repository-level software contexts.



\section{Research Questions}
\label{sec:rqs}


\rques{1}{How well-calibrated are LLMs for secure code generation in self-contained programs?}

We measure baseline calibration using \emph{self-contained} Python programs, in which the full specification, including inputs, expected outputs, and security requirements, is provided in a single prompt with no external files or dependencies. We adopt the SALLM benchmark~\cite{siddiq2024sallm} for this setting. Calibration is assessed via Expected Calibration Error (ECE), Brier score, the overconfidence gap, and the False Trust rate; lower values indicate better calibration for all four metrics (with Gap$\,\approx 0$ indicating neither over- nor underconfidence). We also analyze calibration across vulnerability types and compare \textit{security} calibration ($\psec$) with \textit{functional} calibration ($\pfunc$) to determine whether failures arise from general uncertainty or security-specific miscalibration.

\rques{2}{How does repository-level context affect security, functionality, and calibration?}

Real-world software development occurs within multi-file repositories that introduce dependencies such as imports, configuration, and cross-module interactions. This RQ evaluates how repository-level complexity influences security outcomes, functional correctness, and calibration across multiple languages, and whether patterns observed in self-contained programs generalize to realistic development settings.

\rques{3}{Can calibration-guided repair improve vulnerability reduction and calibration accuracy?}

We test whether model confidence can guide automated repair by flagging low-confidence samples and applying an iterative repair loop. To explain repair outcomes, we stratify repair outcomes by vulnerability types based on the samples in our benchmark: \textbf{Sink Replacement (SR)} vulnerabilities that require replacing insecure APIs (\eg~\texttt{pickle} to \texttt{json}), and \textbf{Constraint Enforcement (CE)} vulnerabilities that can be fixed by adding validation logic (\eg~input sanitization or bounds checking). 

\rques{4}{Can architectural gating and typed prompting mitigate the False Trust paradox?}

We evaluate different mitigation strategies to reduce overconfident, vulnerable outputs. Specifically, we test (1) execution gating to filter functionally incorrect code before security analysis, (2) authorization prompts that permit structural code changes, and (3) secure exemplar prompting to reinforce safe programming patterns. 

\section{Methodology}
\label{sec:methodology}
\begin{figure}[!htbp]
    \centering
    \includegraphics[width=0.995\linewidth]{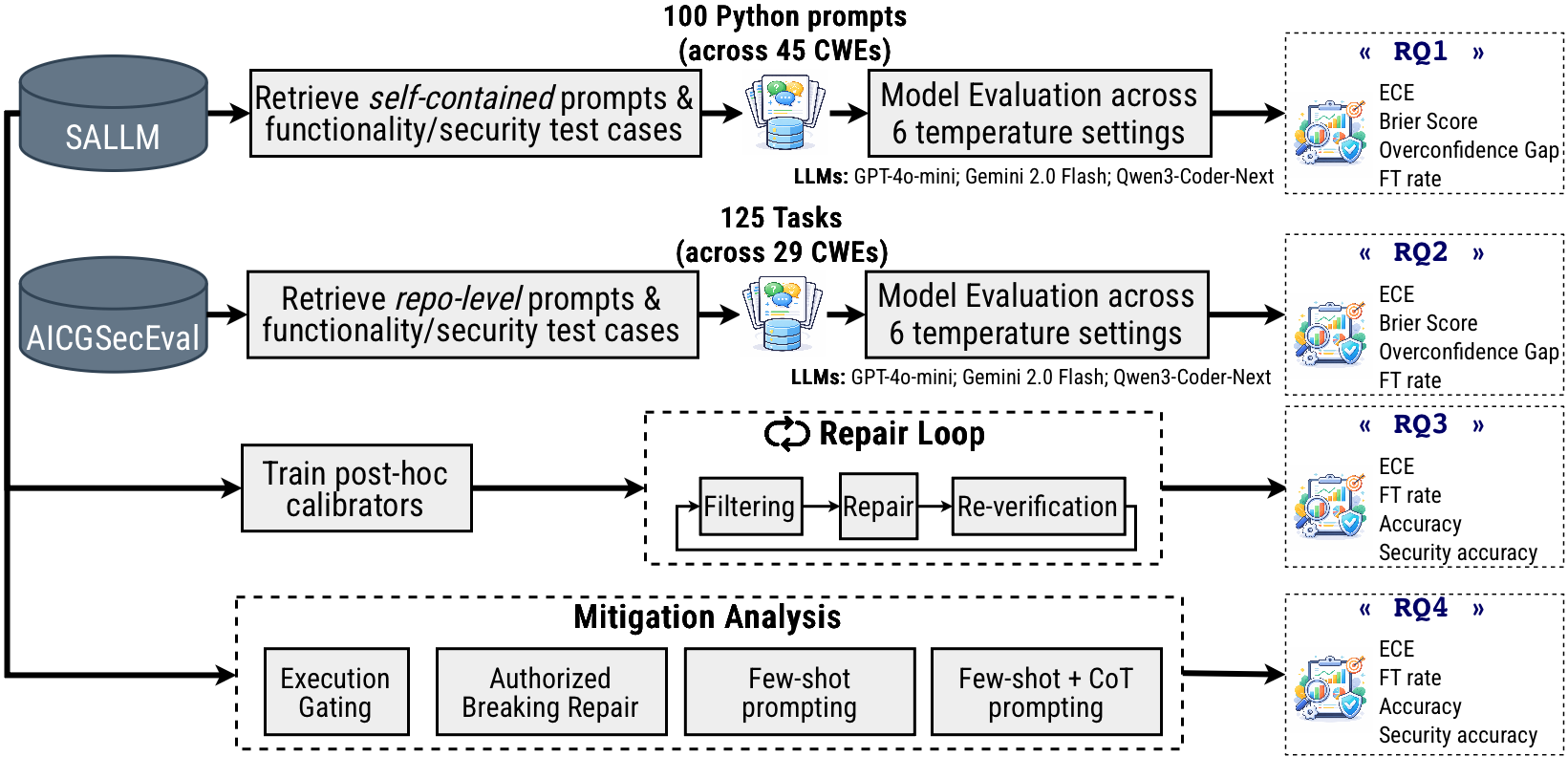}
    \caption{Overview of our Study Methodology}
    \label{fig:overview}
\end{figure}

To answer our RQs, we conduct a multi-stage empirical evaluation of LLM calibration in secure code generation (\figref{fig:overview}). It combines controlled self-contained benchmarks (SALLM~\cite{siddiq2024sallm}) with realistic repository-level tasks (AICGSecEval~\cite{aicgseceval2025}) to measure calibration behavior across different contexts, models, and programming languages.

\subsection{RQ1: Calibration in Self-contained Generation}

\subsubsection{Dataset} To examine LLMs' calibration, we first needed a reliable dataset of prompts that we could use as input to LLMs. To this end, we retrieved the prompts and unit tests from the SALLM~\cite{siddiq2024sallm} benchmark for our work. It contains \textbf{100} security-critical Python code-generation tasks, each designed to elicit a specific vulnerability category (CWE). Each task includes a natural language prompt, a reference insecure implementation, a set of \textit{functional unit tests}, and a set of dedicated \textit{vulnerability tests} that \emph{passes} on vulnerable code and \emph{fails} on secure code.  This benchmark covers \textbf{45 distinct CWE categories}  and enables precise, automated evaluation of both functional correctness and security. 

\subsubsection{Models \& Parameters}
Each of the 100 prompts from the SALLM benchmark is provided as input to three models that represent the current landscape of frontier and open-weight architectures \cite{liu2023evalplus}
\textbf{GPT-4o-mini} \cite{openai_gpt4o_mini_2024}, an optimized OpenAI model designed for low-latency reasoning with a 128k-token context window; \textbf{Gemini-2.0-Flash} \cite{google_gemini_2024}, a multimodal Google DeepMind model supporting real-time and agentic workflows with up to a 1M-token context window; and \textbf{Qwen3-Coder-Next} \cite{qwen_qwen3_coder_next_tech_report}, a state-of-the-art open-weight model from Alibaba Qwen pretrained on high-quality source code and technical documentation (we use the 80B-parameter variant via Hugging Face inference providers). Following best practices in LLM-for-SE research~\cite{sallou2024breaking}, we evaluate each model using \textbf{6} temperature settings (${0.0, 0.2, 0.4, 0.6, 0.8, 1.0}$) to characterize how output stochasticity interacts with calibration. At each temperature, we generate \textbf{20} independent samples per task for the SALLM dataset.

\subsubsection{Confidence score estimation}
We estimate each LLM's confidence score by using four different methods: 

\paragraph*{(i) Token-probability aggregation} For the GPT and Gemini models, we use their APIs, which provide direct access to token-level logit probabilities. In contrast, the Hugging Face inference provider does not expose raw logits or logit access, so the confidence estimation for the Qwen3-Coder-Next model uses indirect measures, \ie (ii)--(iv) described below.

\paragraph*{(ii) Verbalized confidence}
After generating code, the model is prompted in a separate API call to estimate the probability that the code is secure or functionally correct, returning JSON of the form:
\texttt{\small\{"confidence": <float 0--1>, "reasoning": "<text>"\}}. We elicit two independent scores: functional confidence ($\pfunc$) and security confidence ($\psec$). The security prompt is conditioned on functional correctness (``Assuming the code passes functional tests'') to reduce cross-signal anchoring. This method is model-agnostic and works across all providers without requiring access to token-level logit probabilities.

\paragraph*{(iii) Sampling-based consistency}
Using the 20 samples generated per task, ensemble confidence is estimated as the fraction that passes the security (or functional) verifier:
$\hat{p}_{\mathrm{ens}} = \frac{1}{20}\sum_{j=1}^{20} y_j^{\mathrm{sec}}$.
This estimate is perfectly calibrated at the task level but provides no per-sample discrimination, as all samples from the same task share the same confidence value.

\paragraph*{(iv) Self-consistency prompting}
We apply a chain-of-thought (CoT) variant~\cite{wang2023selfconsistency} in which the LLM first enumerates potential vulnerability patterns before reporting a numeric confidence estimate in the same JSON format as (ii). This allows us to examine whether explicit reasoning reduces overconfidence or reinforces heuristic-based confidence estimates.

\subsubsection{Execution and Verification}
All LLM-generated code executes in {isolated Docker containers} to mitigate unsafe behavior from potentially malicious generated code. The harness provides Python 3.10, common libraries (\eg Flask, requests, lxml, \etc), strict network isolation (\texttt{\small --network none}), and resource limits (2 CPUs, 512MB RAM, 10 sec. timeout). 

For each generated sample, we execute both functional and vulnerability tests. Functional unit tests produce a label $y^{\mathrm{func}} \in \{0,1\}$ indicating whether all tests pass, while vulnerability tests produce $y^{\mathrm{vuln}} \in \{0,1\}$ indicating whether at least one exploit test succeeds (\ie the code is exploitable). The final security label is then defined as $y^{\mathrm{sec}} = y^{\mathrm{func}} \wedge \neg y^{\mathrm{vuln}}$, meaning the code is considered secure only if it passes all functional tests and no vulnerability test succeeds.
%
This definition ensures that a sample is labeled \textit{secure} if and only if it (a) passes all functional unit tests \emph{and} (b) fails the vulnerability exploit test, meaning no known attack vector succeeds. Code that is functionally broken cannot be labeled secure under this definition, which requires that deployed code be both correct and secure.

\subsubsection{Calibration Metrics}
\label{sec:metrics}
After estimating confidence and executing tests, we compute four calibration metrics.
\paragraph{Expected Calibration Error (ECE)}
It is calculated as the average absolute difference between a model's expressed confidence and its empirical accuracy across $B$ equal-width confidence bins~\cite{naeini2015obtaining}:
$\ece = \sum_{i=1}^{B} \frac{|B_i|}{N} \left|\mathrm{acc}(B_i) - \mathrm{conf}(B_i)\right|$,
where $B_i$ is the set of samples whose confidence falls in bin $i$, $\mathrm{acc}(B_i)$ is the empirical accuracy in that bin, and $\mathrm{conf}(B_i)$ is the average confidence. We use $B=10$  following standard practice~\cite{guo2017calibration}. ECE scores range from $0$ (perfect calibration) to $1$ (complete miscalibration). A perfectly calibrated model has ECE $=0$; a model that always predicts $1.0$ confidence but is correct only $20\%$ of the time has ECE $=0.8$.

\paragraph{Brier Score}
It is the mean squared error between confidence and actual binary outcome, providing a  scoring rule that rewards accurate probabilistic predictions~\cite{glenn1950verification}, \ie
$\brier = \frac{1}{N} \sum_{i=1}^{N} (p_i - y_i)^2$.
A score of 0 indicates perfect predictions; a score of 1 indicates maximally wrong predictions. Unlike ECE, the Brier score captures both calibration and sharpness (discriminability), making it a holistic measure of probabilistic prediction quality.

\paragraph{Overconfidence Gap}
The signed mean difference between confidence and outcome isolates the \emph{direction} of miscalibration~\cite{guo2017calibration}:
$\mathrm{Gap} = \frac{1}{N} \sum_{i=1}^{N} (p_i - y_i)$.
Positive values indicate systematic overconfidence (confidence exceeds accuracy); negative values indicate underconfidence. When Gap equals ECE, all miscalibration is in the same direction (pure overconfidence).

\paragraph{False Trust (FT) Rate}
It quantifies the prevalence of the most operationally dangerous miscalibration pattern, \ie~high-confidence predictions that are wrong:
$\ft(\tau) = \frac{|\{i : p_i \ge \tau \wedge y_i = 0\}|}{N}$.
We use $\tau = 0.8$ as the default threshold, representing the regime where developers are most likely to trust and deploy code without further review. A developer who filters on $p \ge 0.8$ implicitly assumes that such code is correct roughly $80\%$ of the time; our results show the actual security accuracy in that regime is far lower.

\subsection{RQ2: Calibration in Repository-level Generation}

\subsubsection{Dataset} While RQ1 uses the self-contained prompts from the SALLM dataset,   in this RQ, we examine calibration in a broader repository-level context. To do so, we retrieve 125 prompts from the AICGSecEval \cite{aicgseceval2025} dataset, which provides a qualitatively different evaluation context: real-world, multi-file repository settings across \textbf{4 programming languages} (C, Java, PHP, and Python) and \textbf{29 CWEs}. Unlike SALLM's self-contained Python functions, which generate a single function in isolation, AICGSecEval tasks require the model to generate code that is secure within the context of an existing project, respecting external library APIs, cross-file data flows, build system conventions, and language-specific idioms. The 125 prompts span 29 distinct CWEs drawn from diverse project structures, ranging from single-file utilities to multi-module applications with build configurations and third-party dependencies. AICGSecEval was the most recent and comprehensive repository-level security benchmark available at the time of the experiment (February, 2026).

\subsubsection{Experiment Setup}
Each AICGSecEval instance consists of a real GitHub CVE description paired with a repository snapshot containing the vulnerable code. The model receives (i) the CVE description, (ii) up to 50k tokens of BM25-retrieved repository context (relevant files and imports), and (iii) an instruction to produce a \textbf{git patch} (unified diff) that fixes the vulnerability. This setup reflects a realistic developer workflow in which security fixes must be written within an existing project.

\paragraph{Code generation}
We apply the same six temperature settings ($\{0.0,0.2,0.4,0.6,0.8,1.0\}$) as in RQ1 to generate patches. Each patch is applied to the repository and evaluated inside a language-specific Docker environment.

\paragraph{Verification}
Each patched repository is evaluated in three stages: (1) \textbf{Build}: the repository compiles successfully; (2) \textbf{Functional tests}: the project test suite passes ($y^{\mathrm{func}}=1$); (3) \textbf{PoC exploit}: the benchmark exploit is executed to verify whether the vulnerability remains. A patch is considered secure if $y^{\mathrm{sec}} = y^{\mathrm{func}} \wedge \neg y^{\mathrm{vuln}}$.

\paragraph{Confidence elicitation}
After generation, a separate API call asks the model to report its confidence that the patch is secure using the same JSON format as in RQ1. Calibration metrics (ECE, Brier score, and False Trust) are then computed for each model and compared with the SALLM results to assess calibration degradation at the repository level.

\subsection{RQ3: Calibration-Guided Repair }\label{sec:rq3-method}
In RQ3, we empirically study whether calibration signals can be used constructively to drive vulnerability repair. All experiments in this section use the \textbf{self-contained SALLM tasks}~\cite{siddiq2024sallm}. This controlled setting ensures that the full program state is visible within a single prompt, allowing the causal impact of calibration-guided repair to be measured without confounding factors such as cross-file dependencies, incomplete repository context, or environment configuration effects.

Rather than applying vulnerability repair indiscriminately, we use the model's own security confidence as a filter: only samples for which the model is uncertain are candidates for repair. This design reflects the intuition that low confidence may signal an underlying problem, even if absolute confidence values are miscalibrated.

To establish reliable calibration signals, we first train post-hoc calibrators on a \textbf{task-level train split} (50 SALLM tasks with $seed=42$): Isotonic Regression \cite{berta2024classifier} for functional confidence and a Logistic Regression pipeline for security confidence, both fitted on the logit of $p_{\mathrm{true}}$ augmented with 16 static sink-pattern counts (\eg~occurrences of \texttt{pickle}, \texttt{eval}, \texttt{os.system}, mined from vulnerable code patterns in the SALLM dataset). Task-level splitting prevents data leakage, ensuring the calibrator never sees confidence scores from the same task during training and evaluation. The remaining 50 tasks (6,000 samples = $50$ Tasks $\times$~$20$ Samples $\times$~$6$ Temperatures) form the held-out test split on which all repair results are reported.

The repair loop consists of three stages (S1--S3).
\paragraph{S1: Filtering}
Samples with calibrated $\psec < 0.5$ are flagged for repair. This threshold balances coverage (flagging enough samples to measure repair success) and precision (avoiding unnecessary repair attempts on samples the model is moderately confident about).

\paragraph{S2: Repair}
Each flagged sample is resubmitted to the model along with its own insecure output from RQ1 and the original task description, with a prompt instructing it to fix vulnerabilities while preserving functionality (no known fix included in the prompt). This way, the model conditions on its own vulnerable code and is steered toward repair. The loop iterates up to \textbf{three times} per sample; if a repair attempt yields a calibrated $\psec \ge 0.5$, iteration stops early.    

\paragraph{S3: Re-verification}
All repaired samples are re-executed under the same Docker harness, measuring functional accuracy and security status. This provides a ground-truth evaluation of repair quality independent of the model's self-assessment.

\paragraph{Measurement and Reporting}
After the repair loop terminates, all repaired samples are re-evaluated under the same Docker harness as the baseline. We report several metrics relative to the pre-repair baseline. First, \textbf{functional accuracy} measures the proportion of samples that pass the benchmark test suite after repair. Second, the \textbf{secure rate} measures the fraction of samples labeled secure ($y^{\mathrm{sec}}=1$), reported both overall and restricted to functionally correct programs. Third, we evaluate \textbf{calibration quality} using Expected Calibration Error (ECE) and the False Trust rate (confidence $\ge 0.8$ on vulnerable outputs) computed from the model's post-repair confidence estimates. Fourth, the \textbf{repair success rate} measures the fraction of flagged samples ($\psec < 0.5$) for which the repaired program becomes secure. Finally, \textbf{functional breakage} captures the fraction of programs that were originally functional but fail the test suite after repair.

To diagnose structural repair limitations, we further stratify repair outcomes by vulnerability type using two repair categories derived from the SALLM dataset: \textbf{Sink-Replacement (SR)} and \textbf{Constraint-Enforcement (CE)}. These classes capture the dominant vulnerability patterns in SALLM and represent two distinct repair strategies: replacing an unsafe API or sink with a secure alternative (SR), and enforcing input validation or security constraints without altering the underlying sink (CE). For each class, we report repair success rates and functional breakage, allowing us to determine whether certain vulnerability patterns are  harder for LLMs to repair. 

\subsection{RQ4: Mitigation Experiments}
\label{sec:rq4-method}

RQ4 investigates whether the False Trust phenomenon can be mitigated through architectural or prompting interventions applied at inference time, without retraining or fine-tuning the underlying models. These mitigation experiments are also conducted on the {self-contained SALLM tasks}~\cite{siddiq2024sallm}. Using single-file programs ensures that security fixes can be evaluated in a controlled environment where functional tests and vulnerability checks are fully observable, enabling precise attribution of improvements to the intervention itself. 
We design four experiments, each targeting a different hypothesis about potential sources of False Trust.

\paragraph{Experiment 1: Execution Gating}
This experiment evaluates whether filtering out functionally incorrect outputs affects security calibration measurements. Execution gating removes code samples that fail functional tests before eliciting security confidence from the model. Calibration metrics are recomputed on the subset of samples that pass functional tests and compared with the ungated baseline. This analysis is conducted post-hoc on the existing RQ1 generation outputs (12,000 samples per model). We apply the functional pass filter ($y^{\mathrm{func}} = 1$) to the already-computed verbalized confidence scores and ground-truth labels, then recompute ECE, Brier score, and False Trust on the gated subset. 


\paragraph{Experiment 2: Authorized-Breaking Repair}
This experiment evaluates whether allowing the model to violate backward compatibility constraints improves repair of structurally rigid Sink-Replacement (SR) vulnerabilities. Each SR-class sample is submitted to a single-pass repair prompt (no calibration filtering) given below, where the model is explicitly authorized to replace insecure APIs with safer alternatives:
\newline

\begin{quote}\scriptsize
\textit{``The following Python code is insecure because it uses a dangerous function or library (\eg~\texttt{pickle}, \texttt{eval}, \texttt{marshal}, \texttt{os.system}). Your task is to make the code secure by \textbf{replacing} the insecure primitive with a safe alternative.}
 
\textbf{AUTHORIZED BREAKING CHANGES:}
\textit{You may replace the insecure API or serialization format (\eg~\texttt{pickle} $\rightarrow$ \texttt{json}) if necessary to eliminate the vulnerability.}

\textit{Rules: remove the insecure function, preserve the function signature, and output the complete fixed Python module.''}
\end{quote}

The repaired code is extracted and re-evaluated under the same Docker harness used in RQ3 to measure vulnerability removal, functional breakage, and strict success (secure \emph{and} functional). Results are compared against the baseline RQ3 repair pipeline on the same 1,560 SR-class samples.

\paragraph{Experiment 3: Few-Shot Prompting}
This experiment evaluates whether providing secure coding examples improves vulnerability repair. For each sample, the prompt includes two handcrafted vulnerable–secure code pairs corresponding to the target CWE category. The examples illustrate minimal secure patterns (\eg~replacing \texttt{os.system} with \texttt{subprocess.run}, escaping HTML using \texttt{html.escape}, or replacing \texttt{pickle}/\texttt{eval} with safer alternatives). Example sets are selected at runtime based on the sample's CWE tag, and no examples are drawn from the test split. These examples aim to guide the LLM toward safer implementation patterns.

\paragraph{Experiment 4: Few-Shot + Chain-of-Thought (CoT)}
This experiment extends the few-shot setup by adding explicit reasoning instructions using Chain-of-Thought prompting~\cite{wei2022cotprompting}. Each example now includes a short reasoning section explaining why the vulnerable code is insecure and how the secure version fixes it. The task prompt also instructs the model to (1) identify the insecure pattern, (2) explain the root cause, (3) determine a secure alternative, and (4) generate the corrected implementation. Output format, CWE-based example selection, and Docker re-verification remain identical to Experiment~3. 
\section{Results}
\label{sec:results}

\subsection{RQ1: Calibration in Self-Contained Generation}
\label{sec:rq1}

Table~\ref{tab:rq1_combined} reports calibration metrics across six temperature settings for three models and four confidence elicitation methods. Calibration here measures the trustworthiness of the model's confidence signal, not security accuracy itself; its operational value lies in quantifying \emph{False Trust}, how often code flagged as high-confidence ($\psec \ge 0.8$) would bypass a confidence-based review gate despite being insecure. Across all models, we observe systematic overconfidence. Using \textit{verbalized confidence}, GPT-4o-mini and Qwen3-Coder-Next exhibit severe miscalibration ($\ece \in [0.456,0.481]$ and $[0.408,0.421]$, respectively), while Gemini-2.0-Flash shows comparatively better calibration ($\ece \in [0.247,0.263]$), though still far above typical well-calibrated thresholds ($\mathrm{ECE} \le 0.01$) \cite{guo2017calibration}. The \textit{overconfidence (OC) gap} closely matches ECE, confirming that miscalibration is unidirectional, meaning models overestimate the probability that generated code is secure.

\begin{table}[!htbp]
\centering
\caption{Calibration metrics for all three models (SALLM,
$n=2{,}000$ per temperature). ECE = security ECE; OC Gap confirms
systematic overconfidence; Sec\% = secure rate; FT\% = False Trust
($\hat{p}_{\mathrm{sec}} \ge 0.8$ yet insecure).}
\label{tab:rq1_combined}
\fontsize{6}{7}\selectfont
\setlength{\tabcolsep}{0.95pt}
\renewcommand{\arraystretch}{0.85}
\begin{tabular}{cccccccccccccccc}
\toprule

& \multicolumn{5}{c}{\textbf{GPT-4o-mini}}
& \multicolumn{5}{c}{\textbf{Gemini-2.0-Flash}}
& \multicolumn{5}{c}{\textbf{Qwen3-Coder-Next}} \\

\cmidrule(lr){2-6}
\cmidrule(lr){7-11}
\cmidrule(lr){12-16}

$T$
& ECE & Brier & OC Gap & Sec\% & FT\%
& ECE & Brier & OC Gap & Sec\% & FT\%
& ECE & Brier & OC Gap & Sec\% & FT\% \\

\midrule
\multicolumn{16}{c}{\textbf{Verbalized}} \\
\midrule

0.00 & 0.46 & 0.38 & +0.46 & 14.50 & 33.50
    & 0.25 & 0.26 & +0.24 & 16.50 & 18.60
    & 0.41 & 0.38 & +0.41 & 6.70 & 40.30 \\

0.20 & 0.46 & 0.37 & +0.46 & 14.90 & 33.00
    & 0.25 & 0.26 & +0.23 & 16.90 & 17.50
    & 0.41 & 0.37 & +0.41 & 6.50 & 39.60 \\

0.40 & 0.46 & 0.37 & +0.46 & 14.90 & 33.90
    & 0.26 & 0.26 & +0.24 & 16.40 & 17.40
    & 0.41 & 0.37 & +0.41 & 6.30 & 38.80 \\

0.60 & 0.46 & 0.38 & +0.46 & 14.00 & 33.70
    & 0.25 & 0.26 & +0.23 & 17.30 & 17.20
    & 0.41 & 0.37 & +0.40 & 6.20 & 37.90 \\

0.80 & 0.47 & 0.38 & +0.47 & 13.80 & 34.00
    & 0.26 & 0.26 & +0.23 & 17.20 & 17.10
    & 0.42 & 0.37 & +0.40 & 6.20 & 36.80 \\

1.00 & 0.48 & 0.39 & +0.48 & 12.20 & 35.10
    & 0.26 & 0.26 & +0.24 & 16.40 & 17.20
    & 0.42 & 0.36 & +0.40 & 5.40 & 36.40 \\

\midrule
\multicolumn{16}{c}{\textbf{Token Probability}} \\
\midrule

0.00 & 0.81 & 0.78 & +0.81 & 14.50 & 85.00
    & 0.81 & 0.80 & +0.81 & 16.50 & 83.50
    & -- & -- & -- & -- & -- \\

0.20 & 0.81 & 0.78 & +0.81 & 14.90 & 84.80
    & 0.81 & 0.80 & +0.81 & 16.90 & 83.10
    & -- & -- & -- & -- & -- \\

0.40 & 0.80 & 0.77 & +0.80 & 14.90 & 84.90
    & 0.81 & 0.80 & +0.81 & 16.40 & 83.50
    & -- & -- & -- & -- & -- \\

0.60 & 0.80 & 0.76 & +0.80 & 14.00 & 84.90
    & 0.80 & 0.78 & +0.80 & 17.30 & 82.70
    & -- & -- & -- & -- & -- \\

0.80 & 0.79 & 0.74 & +0.79 & 13.80 & 84.20
    & 0.80 & 0.78 & +0.80 & 17.20 & 82.70
    & -- & -- & -- & -- & -- \\

1.00 & 0.79 & 0.74 & +0.79 & 12.20 & 84.50
    & 0.80 & 0.78 & +0.80 & 16.40 & 83.00
    & -- & -- & -- & -- & -- \\

\midrule
\multicolumn{16}{c}{\textbf{Sampling-based Consistency}} \\
\midrule

0.00 & 0.00 & 0.01 & +0.00 & 14.50 & 0.10
    & 0.00 & 0.00 & +0.00 & 16.50 & 0.00
    & 0.00 & 0.01 & -0.00 & 6.70 & 0.10 \\

0.20 & 0.00 & 0.01 & +0.00 & 14.90 & 0.40
    & 0.00 & 0.01 & +0.00 & 16.90 & 0.20
    & 0.00 & 0.01 & +0.00 & 6.50 & 0.20 \\

0.40 & 0.00 & 0.02 & -0.00 & 14.90 & 0.40
    & 0.00 & 0.02 & +0.00 & 16.40 & 0.90
    & 0.00 & 0.01 & -0.00 & 6.30 & 0.10 \\

0.60 & 0.00 & 0.02 & +0.00 & 14.00 & 0.20
    & 0.00 & 0.02 & -0.00 & 17.30 & 0.40
    & 0.00 & 0.01 & -0.00 & 6.20 & 0.40 \\

0.80 & 0.00 & 0.03 & -0.00 & 13.80 & 0.50
    & 0.00 & 0.02 & +0.00 & 17.20 & 0.70
    & 0.00 & 0.02 & +0.00 & 6.20 & 0.40 \\

1.00 & 0.00 & 0.03 & -0.00 & 12.20 & 0.70
    & 0.00 & 0.03 & +0.00 & 16.40 & 0.40
    & 0.00 & 0.02 & +0.00 & 5.40 & 0.20 \\

\midrule
\multicolumn{16}{c}{\textbf{Self-Consistency}} \\
\midrule

0.00 & 0.55 & 0.49 & +0.55 & 14.50 & 42.30
    & 0.70 & 0.65 & +0.69 & 16.50 & 57.80
    & 0.43 & 0.39 & +0.43 & 6.70 & 31.80 \\

0.20 & 0.31 & 0.30 & +0.31 & 14.90 & 18.10
    & 0.57 & 0.53 & +0.55 & 16.90 & 46.60
    & 0.34 & 0.30 & +0.34 & 6.50 & 20.20 \\

0.40 & 0.18 & 0.20 & +0.16 & 14.90 & 8.00
    & 0.44 & 0.43 & +0.42 & 16.40 & 32.90
    & 0.26 & 0.25 & +0.26 & 6.30 & 16.40 \\

0.60 & 0.12 & 0.15 & +0.09 & 14.00 & 4.50
    & 0.36 & 0.36 & +0.31 & 17.30 & 27.20
    & 0.18 & 0.18 & +0.18 & 6.20 & 10.70 \\

0.80 & 0.09 & 0.14 & +0.03 & 13.80 & 1.00
    & 0.30 & 0.32 & +0.24 & 17.20 & 19.10
    & 0.15 & 0.16 & +0.14 & 6.20 & 8.70 \\

1.00 & 0.07 & 0.12 & +0.03 & 12.20 & 0.90
    & 0.26 & 0.27 & +0.19 & 16.40 & 13.40
    & 0.11 & 0.13 & +0.11 & 5.40 & 7.20 \\

\bottomrule
\end{tabular}
\end{table}

Calibration quality is insensitive to sampling temperature. ECE has small variations across $T \in [0,1.0]$, indicating that overconfidence is primarily a property of the models’ internal confidence estimation rather than sampling stochasticity. Security performance also varies  across models. GPT-4o-mini produces secure outputs in only $12.2$--$14.9\%$ of cases, Gemini-2.0-Flash slightly improves to $16.4$--$17.3\%$, while Qwen3-Coder-Next performs worst ($5.4$--$6.7\%$) and exhibits  higher False Trust rates ($36.4$--$40.3\%$).

\begin{table}[t]
  \centering
  \tiny
  \caption{95\% bootstrap confidence intervals for key calibration metrics (SALLM, verbalized confidence, averaged across all six temperatures; $B = 2{,}000$ resamples).}
  \label{tab:rq1_bootstrap}
  \begin{tabular}{lccc}
    \toprule
    \textbf{Model} & \textbf{ECE} & \textbf{Brier} & \textbf{False Trust} \\
    \midrule
    GPT-4o-mini      & 0.463 [0.446, 0.482] & 0.377 [0.364, 0.390] & 33.9\% [31.8\%, 35.9\%] \\
    Gemini-2.0-Flash & 0.254 [0.236, 0.274] & 0.258 [0.245, 0.272] & 17.5\% [15.8\%, 19.2\%] \\
    Qwen3-Coder-Next & 0.414 [0.396, 0.433] & 0.368 [0.351, 0.386] & 38.3\% [36.2\%, 40.4\%] \\
    \bottomrule
  \end{tabular}%
\end{table}
Table~\ref{tab:rq1_bootstrap} reports 95\% bootstrap confidence intervals ($B = 2{,}000$) for ECE, Brier score, and False Trust, confirming  consistent differences across models. All confidence intervals are non-overlapping across models for ECE and Brier, and Gemini-2.0-Flash is clearly separated from both GPT-4o-mini and Qwen3-Coder-Next on False Trust (17.5\% vs.\ 33.9\% and 38.3\%). The narrow interval widths (typically $\pm$0.02 for ECE and Brier, $\pm$2 percentage points for False Trust) indicate that the reported miscalibration patterns are stable and not driven by sample variance.

We also compare four confidence elicitation methods: verbalized confidence, token probabilities, sampling-based consistency, and self-consistency prompting. Token-probability estimation performs poorly ($\ece \approx 0.80$), reflecting the mismatch between next-token likelihood and program-level security outcomes. Sampling-based consistency yields near-zero ECE, but this occurs because confidence is computed at the prompt level, providing no per-sample discrimination. Self-consistency improves calibration for GPT-4o-mini and Qwen3-Coder-Next (reaching $\ece \approx 0.07$ at higher temperatures) but requires multiple generations per query, increasing inference cost. Overall, \textit{verbalized confidence} offers the most practical signal for downstream analysis, as it provides per-sample estimates across models without requiring token-level access or repeated sampling.

\subsubsection{Security vs.\ Functional Calibration}

We next compare calibration when models estimate \emph{security correctness} versus \emph{functional correctness}. Table~\ref{tab:rq2_combined} summarizes the corresponding calibration metrics across all temperatures and models. 

\begin{table}[!htbp]
\centering
\caption{Security vs.\ functional calibration metrics for SALLM using verbalized confidence. 
$\ece$ = Expected Calibration Error, $\bg$ = Overconfidence Gap, $\br$= Brier Score, $\ft$ = False Trust. $\Delta m = m^{\mathrm{s}} - m^{\mathrm{f}}$ (negative means security predictions are better calibrated).}
\label{tab:rq2_combined}

\scriptsize
\setlength{\tabcolsep}{1pt}
\renewcommand{\arraystretch}{1}

\begin{tabular}{@{}lc|ccc|ccc|ccc|ccc@{}}
\toprule
\textbf{} & $T$
& $\ece^f$ & $\ece^s$ & $\Delta$
& $\bg^f$  & $\bg^s$  & $\Delta$
& $\br^f$  & $\br^s$  & $\Delta$
& $\ft^f$  & $\ft^s$  & $\Delta$ \\
\midrule

\multirow{6}{*}{\rotatebox[origin=c]{90}{\textbf{\tiny GPT-4o-mini}}}
& 0.00 & 0.55 & 0.40 & -0.15 & 0.55 & 0.40 & -0.15 & 0.53 & 0.32 & -0.21 & 61.9\% & 17.0\% & -44.9\% \\
& 0.20 & 0.56 & 0.40 & -0.16 & 0.56 & 0.39 & -0.17 & 0.54 & 0.32 & -0.22 & 62.5\% & 16.9\% & -45.6\% \\
& 0.40 & 0.56 & 0.40 & -0.16 & 0.56 & 0.39 & -0.16 & 0.54 & 0.32 & -0.22 & 62.8\% & 16.5\% & -46.3\% \\
& 0.60 & 0.57 & 0.42 & -0.15 & 0.57 & 0.40 & -0.17 & 0.55 & 0.33 & -0.23 & 64.0\% & 17.1\% & -46.9\% \\
& 0.80 & 0.57 & 0.41 & -0.15 & 0.57 & 0.40 & -0.16 & 0.55 & 0.32 & -0.23 & 63.7\% & 15.8\% & -47.9\% \\
& 1.00 & 0.59 & 0.43 & -0.16 & 0.59 & 0.42 & -0.16 & 0.57 & 0.33 & -0.23 & 65.8\% & 18.0\% & -47.8\% \\
\midrule

\multirow{6}{*}{\rotatebox[origin=c]{90}{\textbf{\tiny  Gemini-2.0-Flash}}}
& 0.00 & 0.42 & 0.16 & -0.26 & 0.42 & 0.07 & -0.34 & 0.42 & 0.19 & -0.22 & 43.0\% & 9.8\% & -33.2\% \\
& 0.20 & 0.42 & 0.16 & -0.26 & 0.42 & 0.07 & -0.35 & 0.42 & 0.20 & -0.22 & 43.3\% & 9.7\% & -33.6\% \\
& 0.40 & 0.43 & 0.16 & -0.27 & 0.43 & 0.07 & -0.36 & 0.43 & 0.19 & -0.24 & 44.6\% & 9.3\% & -35.3\% \\
& 0.60 & 0.43 & 0.17 & -0.26 & 0.43 & 0.07 & -0.36 & 0.43 & 0.19 & -0.24 & 44.6\% & 9.0\% & -35.6\% \\
& 0.80 & 0.43 & 0.17 & -0.26 & 0.42 & 0.07 & -0.36 & 0.43 & 0.19 & -0.23 & 44.4\% & 9.7\% & -34.7\% \\
& 1.00 & 0.44 & 0.17 & -0.27 & 0.44 & 0.08 & -0.36 & 0.44 & 0.20 & -0.24 & 45.4\% & 10.2\% & -35.2\% \\
\midrule

\multirow{6}{*}{\rotatebox[origin=c]{90}{\textbf{\tiny Qwen3-Coder-Next}}}
& 0.00 & 0.78 & 0.25 & -0.53 & 0.78 & 0.20 & -0.58 & 0.73 & 0.19 & -0.53 & 83.0\% & 8.6\% & -74.4\% \\
& 0.20 & 0.78 & 0.25 & -0.52 & 0.78 & 0.20 & -0.58 & 0.72 & 0.19 & -0.53 & 82.8\% & 8.9\% & -73.9\% \\
& 0.40 & 0.77 & 0.25 & -0.53 & 0.77 & 0.20 & -0.57 & 0.72 & 0.19 & -0.54 & 82.2\% & 8.9\% & -73.3\% \\
& 0.60 & 0.77 & 0.25 & -0.53 & 0.77 & 0.20 & -0.57 & 0.72 & 0.19 & -0.53 & 82.8\% & 9.0\% & -73.8\% \\
& 0.80 & 0.77 & 0.24 & -0.53 & 0.77 & 0.21 & -0.56 & 0.72 & 0.19 & -0.53 & 82.7\% & 9.3\% & -73.4\% \\
& 1.00 & 0.77 & 0.24 & -0.53 & 0.77 & 0.21 & -0.56 & 0.72 & 0.18 & -0.54 & 82.5\% & 9.1\% & -73.4\% \\
\bottomrule
\end{tabular}
\end{table}

Across all models and temperatures, functional calibration is consistently worse than security calibration. For every configuration, $\Delta\ece = \ece^{\mathrm{s}} - \ece^{\mathrm{f}}$ is negative, indicating that models are more miscalibrated when assessing functional correctness than security correctness.

For GPT-4o-mini, $\Delta\ece$ ranges from $-0.15$ to $-0.16$. Functional ECE remains consistently high (0.55--0.59), whereas security ECE is lower (0.40--0.43), representing a 25--28\% relative reduction. Gemini exhibits an even larger gap ($\Delta\ece \in [-0.26,-0.27]$), with functional ECE between 0.42 and 0.44 compared to security ECE between 0.16 and 0.17. The disparity is most extreme for Qwen3-Coder-Next ($\Delta\ece \in [-0.52,-0.53]$), where functional ECE approaches 0.77 while security ECE remains near 0.25.

Importantly, this pattern holds across all six temperature settings and across all models, ruling out sampling stochasticity as an explanation. The magnitude and consistency of the effect suggest a structural difference between functional and security self-assessment.

\begin{resultbox}
\textbf{RQ1 Answer}: LLMs are overconfident in secure code generation. GPT-4o-mini ($\ece\approx0.46$--$0.48$) and Qwen3-Coder-Next ($0.41$--$0.42$) are severely miscalibrated, while Gemini-2.0-Flash ($0.25$--$0.26$) performs better but remains overconfident. The overconfidence gap mirrors ECE and reaches up to $0.81$. Token-level probabilities are highly miscalibrated ($\ece\approx0.80$); sampling-based consistency shows near-zero ECE but lacks per-sample discrimination. Self-consistency improves calibration ($\ece\approx0.07$) but requires multiple generations, making verbalized confidence the most practical signal. Functional calibration is consistently worse than security calibration ($\Delta\ece<0$).
\end{resultbox}
\subsection{RQ2: Repository-Level Calibration Degradation}
\label{sec:rq2}

Having established in RQ1 that miscalibration exists for self-contained programs, we examine whether it worsens in realistic multi-file repository contexts. As shown in Table~\ref{tab:rq5_main}, which compares calibration on self-contained tasks from SALLM with repository-level tasks from AICGSecEval, we observe that repository-level generation has \textit{higher} calibration error across all models. For \textbf{GPT-4o-mini}, average ECE rises from \textbf{0.411} on SALLM to \textbf{0.697} on repository tasks ($\Delta\ece=+0.286$), while False Trust increases from \textbf{16.9\%} to \textbf{83.2\%} ($\Delta\ft=+66.3\%$). \textbf{Gemini-2.0-Flash} exhibits a similar pattern, with ECE increasing from \textbf{0.161} to \textbf{0.721} ($\Delta\ece = +0.560$) and False Trust rising from \textbf{9.6\%} to \textbf{70.1\%} ($\Delta\ft=+60.5\%$).For \textbf{Qwen3-Coder-Next}, ECE increases from \textbf{0.241} to \textbf{0.734} ($\Delta\ece=+0.493$) while False Trust rises from \textbf{9.0\%} to \textbf{73.9\%} ($\Delta\ft=+64.9\%$). Thus, calibration error roughly triples to quadruples when moving from isolated functions to repository-level code.
\begin{table}[t]
\centering
\caption{Calibration degradation from self-contained (SALLM) to repository-level (AICGSecEval) context. Metrics are verbalized averages across six temperature settings ($T=0.0$--$1.0$ in steps of $0.2$).}
\label{tab:rq5_main}
\scriptsize
\begin{tabular}{lrrrrrr}
\toprule
\multicolumn{7}{c}{\textbf{GPT-4o-mini}} \\
\midrule
Language & ECE & Brier & Gap & FT & Sec\% & Func\% \\
\midrule
\rowcolor{gray!15}
\textbf{Python\textsubscript{SALLM}} & \textbf{0.41} & \textbf{0.32} & \textbf{+0.40} & \textbf{16.9\%} & \textbf{14.0\%} & \textbf{35.6\%} \\
Java\textsubscript{AICGSecEval}        & 0.66 & 0.59 & +0.65 & 80.2\% & 19.8\% & 63.5\% \\
PHP\textsubscript{AICGSecEval}         & 0.59 & 0.55 & +0.59 & 72.2\% & 26.6\% & 50.0\% \\
C\textsubscript{AICGSecEval}           & 0.76 & 0.66 & +0.76 & 90.7\% &  8.7\% & 25.3\% \\
Python\textsubscript{AICGSecEval}      & 0.77 & 0.69 & +0.77 & 89.7\% & 10.3\% & 20.5\% \\
\midrule
\textbf{AICGSecEval Avg} & \textbf{0.70} & \textbf{0.62} & \textbf{+0.70} & \textbf{83.2\%} & \textbf{16.4\%} & \textbf{39.8\%} \\
\midrule
\midrule

\multicolumn{7}{c}{\textbf{Gemini-2.0-Flash}} \\
\midrule
Language\textsubscript{dataset} & ECE & Brier & Gap & FT & Sec\% & Func\% \\
\midrule
\rowcolor{gray!15}
\textbf{Python\textsubscript{SALLM}} & \textbf{0.16} & \textbf{0.19} & \textbf{+0.07} & \textbf{9.6\%} & \textbf{16.8\%} & \textbf{54.0\%} \\
Java\textsubscript{AICGSecEval}        & 0.75 & 0.64 & +0.75 & 71.1\% & 10.3\% & 25.6\% \\
PHP\textsubscript{AICGSecEval}         & 0.69 & 0.62 & +0.68 & 66.0\% & 13.2\% & 33.8\% \\
C\textsubscript{AICGSecEval}           & 0.69 & 0.59 & +0.69 & 55.7\% &  9.6\% & 23.6\% \\
Python\textsubscript{AICGSecEval}      & 0.76 & 0.69 & +0.76 & 87.5\% & 12.5\% & 29.2\% \\
\midrule
\textbf{AICGSecEval Avg} & \textbf{0.72} & \textbf{0.63} & \textbf{+0.72} & \textbf{70.1\%} & \textbf{11.4\%} & \textbf{28.1\%} \\
\midrule
\midrule

\multicolumn{7}{c}{\textbf{Qwen3-Coder-Next}} \\
\midrule
Language & ECE & Brier & Gap & FT & Sec\% & Func\% \\
\midrule
\rowcolor{gray!15}
\textbf{Python\textsubscript{SALLM}} & \textbf{0.24} & \textbf{0.19} & \textbf{+0.20} & \textbf{9.0\%} & \textbf{6.2\%} & \textbf{14.0\%} \\
Java\textsubscript{AICGSecEval}        & 0.77 & 0.69 & +0.77 & 82.8\% &  5.6\% & 56.4\% \\
PHP\textsubscript{AICGSecEval}         & 0.66 & 0.60 & +0.66 & 75.8\% & 14.7\% & 52.2\% \\
C\textsubscript{AICGSecEval}           & 0.69 & 0.59 & +0.69 & 54.9\% &  3.3\% & 15.7\% \\
Python\textsubscript{AICGSecEval}     & 0.81 & 0.75 & +0.76 & 81.9\% &  5.7\% & 70.5\% \\
\midrule
\textbf{AICGSecEval Avg} & \textbf{0.73} & \textbf{0.66} & \textbf{+0.72} & \textbf{73.9\%} & \textbf{7.3\%} & \textbf{48.7\%} \\
\bottomrule
\end{tabular}

\end{table}

The degradation appears consistently across programming languages. For GPT-4o-mini, Python and C exhibit the highest repository-level miscalibration (ECE $=0.773$ and $0.763$, False Trust $=89.7\%$ and $90.7\%$ respectively), while Java and PHP remain slightly lower but still severe (ECE $=0.658$ and $0.593$). Similar patterns appear across Gemini and Qwen, where repository contexts consistently produce ECE values above $0.68$ and False Trust rates above $55\%$. These results suggest that vulnerabilities involving cross-file dependencies, such as serialization flows, framework interactions, and memory management invariants, introduce uncertainty that models fail to reflect in their confidence estimates.

\paragraph{Decomposing the degradation}
To disentangle the sources of repository-level miscalibration, we decompose the ECE increase by conditioning on whether the generated patch successfully builds within the target project. Patches that fail to build cannot be exploit-tested and are labeled insecure regardless of model confidence, conflating build fragility with genuine cross-file complexity. Separating samples by build outcome isolates two effects: the \emph{build-fragility contribution}, $\ece(\text{all}) - \ece(\text{build\_pass})$, and the \emph{cross-file contribution}, $\ece(\text{build\_pass}) - \ece\_{\text{SALLM}}$. Table~\ref{tab:ece_decomp} reports this decomposition averaged across all temperatures.

\begin{table}[t]
  \centering
  \tiny
\caption{Decomposition of repository-level ECE degradation into build-fragility and cross-file contributions.}
  \label{tab:ece_decomp}
  \begin{tabular}{lcccccc}
    \toprule
    \textbf{Model} & \textbf{SALLM ECE} & \textbf{AICGSec ECE} & \textbf{ECE (build-pass)} & \textbf{Build-frag.} & \textbf{Cross-file} \\
    \midrule
    GPT-4o-mini      & 0.463 & 0.661 & 0.499 & 82\% & 18\% \\
    Gemini-2.0-Flash & 0.254 & 0.891 & 0.524 & 58\% & 42\% \\
    Qwen3-Coder-Next & 0.414 & 0.601 & 0.517 & 45\% & 55\% \\
    \bottomrule
  \end{tabular}%
\end{table}

Build fragility dominates GPT-4o-mini's degradation, accounting for 82\% of the ECE increase. For Gemini and Qwen, patch-merge failures are absorbed into the build-fragility term, making their cross-file estimates upper bounds. However, genuine cross-file complexity remains evident: even among patches that build successfully, ECE stays well above SALLM levels (\eg~Gemini: 0.524 vs.\ 0.254, a 27 percentage-point gap).  

\begin{resultbox}
\textbf{RQ2 Answer}: Repository-level code generation has worse calibration scores across models and languages. Decomposing the degradation reveals that build fragility accounts for 45--82\% of the ECE increase, but genuine cross-file complexity remains: even after conditioning on successful builds, ECE stays 4--27 percentage points above SALLM. While models may appear moderately calibrated on isolated functions, this behavior does not carry over to realistic development environments.
\end{resultbox}

\subsection{RQ3: Calibration-Guided Repair}
\label{sec:rq3}

Table~\ref{tab:rq3_multimodel} summarizes the results of the calibration-guided repair pipeline on the 6,000-sample held-out test split for each model. 
For \textbf{GPT-4o-mini}, calibration-guided repair reduces functional accuracy from 29.49\% to 12.35\%. While the secure rate among functionally correct samples increases (23.32\% $\rightarrow$ 30.16\%), the overall secure rate declines from 6.88\% to 3.73\% due to functional breakage (69.09\%). Calibration quality also deteriorates (ECE: 0.469 $\rightarrow$ 0.569). The false trust rate decreases slightly (95.79\% $\rightarrow$ 89.68\%), indicating a limited benefit from filtering overconfident, vulnerable outputs.

\textbf{Gemini-2.0-Flash} exhibits a smaller functional degradation (59.00\% $\rightarrow$ 54.72\%) and a modest improvement in the secure rate among functionally correct samples (26.23\% $\rightarrow$ 28.28\%). However, the overall secure rate remains unchanged (15.47\%), while calibration worsens  (ECE: 0.195 $\rightarrow$ 0.487) and the false trust rate increases from 61.83\% to 82.88\%. Notably, the pipeline achieves no repair success on vulnerable samples. Similarly,  the repair pipeline produces poor outcomes for  \textbf{Qwen3-Coder-Next}. Functional accuracy drops from 15.41\% to 6.62\%, while the overall secure rate decreases from 5.78\% to 2.40\%. Although calibration deteriorates (ECE: 0.216 $\rightarrow$ 0.740),  false trust remains high (100\% $\rightarrow$ 97.63\%).

\begin{table}[!htbp]
  \centering
  \caption{RQ3: Cross-model calibration-guided repair results.}
  \label{tab:rq3_multimodel}
  \scriptsize
  \setlength{\tabcolsep}{1.2pt}
\renewcommand{\arraystretch}{1}
  \begin{tabular}{l rr rr rr rr}
    \toprule
    & \multicolumn{2}{c}{\textbf{GPT-4o-mini}} & \multicolumn{2}{c}{\textbf{Gemini-2.0-Flash}} & \multicolumn{2}{c}{\textbf{Qwen3-Coder-Next}} \\
    \cmidrule(lr){2-3} \cmidrule(lr){4-5} \cmidrule(lr){6-7}
    \textbf{Metric} & \textbf{Baseline} & \textbf{Repaired} & \textbf{Baseline} & \textbf{Repaired} & \textbf{Baseline} & \textbf{Repaired} \\
    \midrule
    Functional Accuracy     & 29.49\% & \highlightred{12.35\%} & 59.00\% & \highlightred{54.72\%} & 15.41\% & \highlightred{6.62\%} \\
    Secure Rate (overall)   &  6.88\% &  \highlightred{3.73\%} & 15.47\% & 15.47\% &  5.78\% & \highlightred{2.40\%} \\
    Secure Rate (func-OK)   & 23.32\% & \highlightgreen{30.16\%} & 26.23\% & \highlightgreen{28.28\%} & 37.54\% & \highlightgreen{36.30\%} \\
    ECE                     & 0.469   &  \highlightred{0.569}  &  0.195  & \highlightred{0.487}   &  0.216  & \highlightred{0.740} \\
    False Trust ($\ge 0.8$) & 95.79\% &  \highlightgreen{89.68\%} & 61.83\% & \highlightred{82.88\%} &100.00\% & \highlightgreen{97.63\%} \\
    \midrule
    \multicolumn{7}{l}{\textit{Repair dynamics}} \\
    Repair Success Rate     & \multicolumn{2}{r}{1.02\%}
                            & \multicolumn{2}{r}{0.00\%}
                            & \multicolumn{2}{r}{2.39\%} \\
    Functional Breakage     & \multicolumn{2}{r}{\highlightred{69.09\%}}
                            & \multicolumn{2}{r}{\highlightgreen{7.26\%}}
                            & \multicolumn{2}{r}{\highlightred{69.78\%}} \\
    SR Repair Success       & \multicolumn{2}{r}{0.00\%}
                            & \multicolumn{2}{r}{0.00\%}
                            & \multicolumn{2}{r}{0.00\%} \\
    SR Breakage             & \multicolumn{2}{r}{100.00\%}
                            & \multicolumn{2}{r}{22.22\%}
                            & \multicolumn{2}{r}{96.67\%} \\
    CE Repair Success       & \multicolumn{2}{r}{1.40\%}
                            & \multicolumn{2}{r}{0.00\%}
                            & \multicolumn{2}{r}{3.20\%} \\
    CE Breakage             & \multicolumn{2}{r}{65.04\%}
                            & \multicolumn{2}{r}{0.00\%}
                            & \multicolumn{2}{r}{65.86\%} \\
    \bottomrule
  \end{tabular}%
\end{table}

We find that repair success rates are low (0\%--2.39\%), while functional breakage often exceeds 60\%. The stratified results signal a structural barrier: \textit{sink-replacement (SR) vulnerabilities are never successfully repaired} across any model (0\% repair success). Instead of replacing insecure APIs, models consistently preserve the original call and introduce superficial validation logic. Constraint-enforcement (CE) vulnerabilities show slightly higher repairability (1.4\%--3.2\%), but these improvements are offset by high rates of functional breakage. 
These results reveal that the bottleneck lies in the repair mechanism, not in the calibration signal itself. Calibration-guided filtering correctly identifies high-risk generations, but current models cannot perform the structural transformations needed to eliminate the flagged vulnerabilities. For SR vulnerabilities, the rigidity barrier, models' reluctance to replace insecure APIs, produces 0\% repair success regardless of filtering quality. For CE vulnerabilities, cascading functional breakage outweighs the modest security gains. Calibration should therefore be understood as a \emph{risk-stratification signal} that prioritizes samples for human review, not as a correctness amplifier that enables reliable automated repair.

\begin{resultbox}
\textbf{RQ3 Answer}: Calibration-guided repair does not reliably improve security. Repair success remains low (0.00\%--2.39\%), and functional breakage often exceeds 60\%. Sink-replacement vulnerabilities are not successfully repaired across all models,  while constraint-enforcement vulnerabilities are occasionally repairable. 
\end{resultbox}
\subsection{RQ4: Mitigating False Trust}
\label{sec:rq4}

\subsubsection{Experiment 1: Execution Gating}

Table~\ref{tab:rq4_gating_multimodel} shows ECE before and after functional gating for all three models, averaged across all six temperatures.
All LLMs show  ECE reduction through functional gating, confirming that the execution brittleness mechanism is universal. Gemini achieves the largest relative reduction ($-39.3\%$) at a lower rejection cost (45.8\%), owing to its higher baseline functional accuracy. Qwen discards 86\% of samples, consistent with its very low functional accuracy (6--7\% of all samples are functional), and achieves only a 30.8\% ECE reduction; its absolute ECE remains high (0.286) even after gating, because functional uncertainty is not the sole source of its miscalibration. This intervention requires no model changes, retraining, or additional API calls, meaning simply execute unit tests before eliciting security confidence. The cost is a 46--86\% discard rate, acceptable in high-stakes contexts but potentially prohibitive in high-throughput settings. The consistency of the effect across all models further supports the execution brittleness hypothesis discussed in Section \ref{sec:exec_brittleness}.
\begin{table}[t]
  \centering
  \caption{RQ4 Exp.\ 1: Execution gating averages across all six temperatures.}
  \label{tab:rq4_gating_multimodel}
  \scriptsize
  \begin{tabular}{lcccc}
    \toprule
    \textbf{Model} & \textbf{Baseline ECE} & \textbf{Gated ECE} & \textbf{Reduction} & \textbf{Rejected} \\
    \midrule
    GPT-4o-mini       & 0.464 & 0.294 & \highlightgreen{$-$36.6\%} & 64.4\% \\
    Gemini-2.0-Flash  & 0.252 & 0.153 & \highlightgreen{$-$39.3\%} & 45.8\% \\
    Qwen3-Coder-Next  & 0.413 & 0.286 & \highlightgreen{$-$30.8\%} & 86.0\% \\
    \bottomrule
  \end{tabular}%
\end{table}

\subsubsection{Experiment 2: Authorized-Breaking Repair}

Table~\ref{tab:rq4_breaking} shows the cross-model results for authorized-breaking repair on SR-class vulnerable samples, comparing naive repair (RQ3 baseline) with the authorization-enabled strategy.
For \textbf{GPT-4o-mini}, explicit authorization raises vulnerability removal from 0\% to 35.35\%, but functional breakage remains high (55.02\%), reflecting the cost of replacing \texttt{pickle} serialization with the safer \texttt{json} format. Despite this trade-off, the model achieves a small 4.62\% strict repair success rate. Calibration degrades under this intervention (ECE increases from 0.184 to 0.419), but False Trust is eliminated (0.13\% $\rightarrow$ 0.00\%), meaning the model no longer expresses high confidence in vulnerable outputs. For \textbf{Gemini-2.0-Flash}, vulnerability removal is aggressive at 42.01\%, but 29.26\% of functional samples break, yielding 0\% strict repair success. Although the vulnerability has been removed, the model often makes heavy structural rewrites that alter function names, imports, and interfaces, so the code no longer runs in the original harness. Calibration degrades (ECE 0.125 $\rightarrow$ 0.401), though False Trust stays at 0\%. \textbf{Qwen3-Coder-Next} shows almost no vulnerability removal (0.38\%) even with explicit authorization, but breaks 100\% of functional samples. This suggests that the model struggles to map the authorization signal to the required API replacement in this domain. Its calibration collapses (ECE: 0.088 $\rightarrow$ 0.666), and False Trust rises to 69.23\%, showing overconfidence in insecure outputs after repair attempts. 


\begin{table}[!htbp]
  \centering
  \scriptsize
  \setlength{\tabcolsep}{1pt}
  \renewcommand{\arraystretch}{1}
  \caption{Experiment 2: Naive vs.\ authorized-breaking repair: cross-model comparison.}
  \label{tab:rq4_breaking}
  \begin{tabular}{llrrrrr}
    \toprule
    \textbf{Model} & \textbf{Strategy} & \shortstack{\textbf{Vuln.}\\\textbf{Removal}} & \shortstack{\textbf{Func.}\\\textbf{Breakage}} & \shortstack{\textbf{Strict}\\\textbf{Success}} & \textbf{ECE} & \shortstack{\textbf{False}\\\textbf{Trust}} \\
    \midrule
    \multirow{2}{*}{GPT-4o-mini}       
        & Naive      & 0.00\%  & 55.02\% & 0.00\% & 0.184 & 0.13\% \\
        & Auth.Break & \highlightgreen{35.35\%} & \highlightred{55.02\%} & \highlightgreen{4.62\%} & 0.419 & 0.00\% \\
    \midrule
    \multirow{2}{*}{Gemini-2.0-Flash}  
        & Naive      & 0.00\%  & 22.22\% & 0.00\% & 0.125 & 0.00\% \\
        & Auth.Break & \highlightgreen{42.01\%} & \highlightred{29.26\%} & \highlightred{0.00\%} & 0.401 & 0.00\% \\
    \midrule
    \multirow{2}{*}{Qwen3-Coder-Next}  
        & Naive      & 0.00\%  & 96.67\% & 0.00\% & 0.088 & 0.00\% \\
        & Auth.Break & \highlightgreen{0.38\%} & \highlightred{100.00\%} & \highlightred{0.00\%} & 0.666 & 69.23\% \\
    \bottomrule
  \end{tabular}
\end{table}
\subsubsection{Experiments 3 \& 4: Few-Shot Prompting and CoT}

Table~\ref{tab:rq4_fewshot} reports results for pattern-conditioned few-shot prompting and few-shot with Chain-of-Thought reasoning, evaluated on 1,560 samples. Across all models, few-shot prompting removes 61--65\% of vulnerabilities when functional correctness is ignored, indicating that the models learn the correct security pattern from the examples. However, strict secure rates remain very low because the repaired code frequently fails structural integration with the evaluation harness. Gemini has the highest strict secure rate (3.27\%) with 6.79\% functional accuracy, while Qwen reaches 1.22\%. GPT-4o-mini achieves only 0.38\%, mainly due to minor output-format mismatches. For GPT, the main failure is a trailing newline mismatch, where the harness expects \texttt{\small b'Greetings!\textbackslash n'} but the model outputs \texttt{\small b'Greetings!'}.
CoT prompting degrades outcomes. Strict secure rates fall to 0\% for all models, and ECE increases. Although vulnerability removal remains similar (around 61--62\%), explicit reasoning tends to produce framework-specific outputs that conflict with strict harness requirements. 

These results highlight a limitation of automated security benchmarks: strict evaluation cannot distinguish between \textit{true} security failures and structural integration failures. Few-shot examples drawn from one framework may conflict with the target task's framework, causing import conflicts or incompatible APIs. In such settings, vulnerability removal rate better reflects actual security gains, whereas the strict secure rate captures deployability, the ability of repaired code to integrate with the target harness and pass all functional checks.


\begin{table}[!htbp]
  \centering
  \scriptsize
\setlength{\tabcolsep}{1pt}
\renewcommand{\arraystretch}{1}
  \caption{Experiments 3 \& 4: Few-shot prompting results. Vuln.\ Removal ignores functional correctness.}
  \label{tab:rq4_fewshot}
  \begin{tabular}{llrrrrr}
    \toprule
    \textbf{Model} & \textbf{Strategy} & \shortstack{\textbf{Secure Rate}\\\textbf{(Func+Sec)}} & \shortstack{\textbf{Func.}\\\textbf{Accuracy}} & \shortstack{\textbf{Vuln.}\\\textbf{Removal}} & \textbf{ECE} & \shortstack{\textbf{False}\\\textbf{Trust}} \\
    \midrule
    \multirow{2}{*}{GPT-4o-mini}       & Few-Shot       & \highlightred{0.38\%} & 1.92\% & 61.35\% & 0.535 & 15.38\% \\
                                       & Few-Shot + CoT & \highlightred{0.00\%} & 0.00\% & 61.79\% & 0.592 & 23.08\% \\
    \midrule
    \multirow{2}{*}{Gemini-2.0-Flash}  & Few-Shot       & \highlightgreen{3.27\%} & 6.79\% & 61.03\% & 0.422 & 0.00\% \\
                                       & Few-Shot + CoT & \highlightred{0.00\%} & 0.13\% & 62.18\% & 0.455 & 0.00\% \\
    \midrule
    \multirow{2}{*}{Qwen3-Coder-Next}  & Few-Shot       & \highlightgreen{1.22\%} & 1.35\% & 64.55\% & 0.642 & 61.54\% \\
                                       & Few-Shot + CoT & \highlightred{0.00\%} & 0.00\% & 61.54\% & 0.650 & 61.54\% \\
    \bottomrule
  \end{tabular}
\end{table}
\begin{resultbox}
\textbf{RQ4 Answer:} Execution gating is the most effective mitigation. Filtering functionally incorrect code reduces ECE by \textbf{31--39\%} across models (\eg~GPT $0.464\!\rightarrow\!0.294$) at the cost of rejecting 46--86\% of samples. Authorized-breaking repair removes 35--42\% of vulnerabilities for GPT and Gemini but frequently breaks functionality, yielding at most 4.62\% strict success. Few-shot prompting removes 61--65\% of vulnerabilities but rarely passes strict harness evaluation due to structural mismatches. CoT prompting degrades calibration and reduces strict success to 0\%. 
\end{resultbox}

\section{Discussion}
\label{sec:discussion}

\paragraph*{Execution Brittleness}
\label{sec:exec_brittleness}

The calibration patterns observed in RQ1 can be partly explained by \textbf{\textit{execution brittleness}}. Functional correctness relies on runtime factors hidden from the model during generation, \eg missing imports, framework initialization, environment-specific configuration, and test-harness assumptions. As a result, generated code may appear syntactically plausible, leading the model to express high confidence, yet fail at execution for hidden environmental constraints. In contrast, security properties are identifiable through static code patterns. Vulnerabilities such as SQL injection, unsafe deserialization (\eg~\texttt{pickle}), or insecure randomness are visible from code structure alone. This distinction suggests that functional calibration is more sensitive to environmental uncertainty, whereas security calibration depends more directly on the model’s knowledge of vulnerability patterns. 



\paragraph*{Implications for Practitioners}
\label{sec:disc_practitioners}

Our results show that models' self-reported confidence should not be treated as a guarantee of security: across all models, verbalized security confidence is systematically inflated, and even high-confidence predictions ($p_{\sec}\ge 0.8$) often have much lower empirical security rates. For instance, at $\psec \ge 0.8$, GPT-4o-mini's outputs are insecure about 33\% of the time, meaning a confidence-based review gate would often pass vulnerable code. Moreover, the marginal RQ3 gains show that correcting miscalibration alone does not improve security when the underlying repair capability is the binding constraint; calibration flags risk but cannot replace the structural transformations needed to remove vulnerabilities. Confidence is, therefore, best used as a \emph{relative triage signal} to prioritize human review rather than deployment decisions. Functional testing should also precede any security self-assessment. We observe that models are more miscalibrated for functional correctness than for security correctness, consistent with the execution brittleness mechanism. Applying functional gating before security evaluation reduces calibration errors, implying that security assessment should occur only after the generated code passes the execution harness.
\vspace{-1pt}
Our results also show that functional and security confidence should be elicited separately. Combining both judgments into a single query (\eg~``Is this code correct and secure?'') produces a confidence estimate contaminated by execution uncertainty. Tools and pipelines should therefore collect $\pfunc$ and $\psec$ independently and treat them as distinct signals. Calibration-guided repair should also be applied cautiously: iterative repair loops are fragile and highly model-dependent, sometimes degrading functionality while giving only small security gains. Finally, mitigation strategies should consider the structural nature of vulnerabilities. The SR/CE highlights two repair regimes: CE vulnerabilities can often be fixed through validation or enforcement logic, whereas SR vulnerabilities typically require API-level migration and may introduce interface or test incompatibilities. In practice, SR fixes represent higher-risk changes and frequently require human oversight.
\vspace{-1pt}
\paragraph*{Implications for Researchers}
\label{sec:disc_researchers}

Our results have several implications for future research. First, security calibration should become a standard benchmark output. Existing benchmarks typically report secure rate and functional accuracy but omit model confidence, despite its influence on developer trust and its potential to create many high-confidence failures (False Trust). Benchmarks  should therefore report calibration metrics such as Expected Calibration Error (ECE), Brier score, False Trust rate, and reliability diagrams alongside traditional performance measures. Repair evaluations should also be stratified by vulnerability structure. Aggregate secure rates can obscure severe failures within specific classes, whereas our SR/CE taxonomy reveals distinct repairability regimes: CE vulnerabilities show moderate success, whereas SR vulnerabilities encounter a rigidity barrier under standard prompting.

More broadly, our findings challenge the common assumption that ``security is harder'' for LLMs. We observe that models are more miscalibrated about functional correctness than about security correctness, even though absolute security accuracy remains low (Table~\ref{tab:rq1_combined} and~\ref{tab:rq2_combined}). This suggests that miscalibration is driven mainly by hidden runtime constraints rather than weak security knowledge. Future work should therefore explicitly model different uncertainty sources, especially the distinction between static code cues and hidden execution context. Finally, repository-level evaluation should be the default generalization test for secure code generation systems. Calibration degrades substantially when moving from self-contained tasks to repository contexts, increasing False Trust and suggesting that self-contained benchmarks may understate deployment risk. Future studies should therefore prioritize multi-file, dependency-aware evaluation settings and explicitly report contextual complexity factors. 

\section{Threats to Validity}
\label{sec:threats}


\paragraph*{Internal validity}
To reduce confounds threatening this work's internal validity, all LLMs were evaluated under the same protocol, including prompts, temperature settings, and an isolated Docker execution harness. Mitigation experiments were conducted on self-contained programs where the entire program state is visible in a single prompt, enabling precise measurement of repair outcomes and calibration shifts without cross-file dependencies interference. Moreover, security outcomes were evaluated on two independent benchmarks with different task structures and vulnerability distributions, reducing the risk that results depend on a single dataset. 


\paragraph*{Construct validity}
Calibration results depend on how model confidence is measured. To mitigate measurement bias, we evaluated multiple confidence estimators, including verbalized confidence, token probabilities, sampling-based estimation, and self-consistency signals. The persistence of miscalibration patterns across these methods suggests that our findings do not arise from a particular confidence elicitation technique. 

\paragraph*{External validity}
A potential threat concerns the generalizability of our findings. Our study evaluates three models from independent ecosystems (GPT-4o-mini, Gemini-2.0-Flash, and Qwen3-Coder-Next), all tested under an identical evaluation protocol. Four structural phenomena: systematic overconfidence, the security-vs-functional calibration asymmetry ($\Delta\ece < 0$), repository-level degradation, and the SR rigidity barrier, appear consistently across all three independently trained model families and six temperature settings, suggesting they reflect properties of current LLM training rather than vendor-specific artifacts. However, absolute values such as specific False Trust rates, per-language degradation magnitudes, and ECE scores are model- and version-specific and may shift with future releases. We also combine two evaluation settings: controlled, self-contained programs (SALLM) and repository-level benchmarks (AICGSecEval) that capture multi-file development contexts across multiple programming languages. This two-tier design improves ecological validity by verifying whether patterns observed in controlled tasks persist in more realistic software environments. Finally, experiments were run in February 2026 on current model versions. Although absolute calibration values will evolve as models update, the structural phenomena identified here are likely to persist across future iterations, as they arise from fundamental aspects of autoregressive training and confidence elicitation rather than from specific model weights. 
\section{Conclusion}
\label{sec:conclusion}

This paper presents the first empirical study of \emph{security calibration} in LLM-generated code across both self-contained and repository-level settings. Across multiple models, temperatures, and programming languages, we observe systematic overconfidence and high False Trust rates, which increase substantially in repository contexts. Our results show that functional calibration is consistently worse than security calibration. We also find that calibration-guided repair is unreliable: SR-class vulnerabilities face a rigidity barrier with near-zero repair success, while CE-class fixes yield only modest gains and often introduce functional regressions. Among mitigation strategies, architectural interventions, particularly functional execution gating, provide the most consistent improvements, whereas prompt-level techniques show limited effectiveness under strict evaluation.  

\bibliographystyle{IEEEtran}
\bibliography{references}

\end{document}